\documentclass{PoS}

\def\to{\rightarrow}

\def\bi{\begin{itemize}}
\def\ei{\end{itemize}}

\def\ta{\tilde a}

\def\tst{\tilde t}

\def\tg{\tilde g}

\def\tw{\widetilde W}
\def\tz{\widetilde Z}
\def\alt{\stackrel{<}{\sim}}
\def\agt{\stackrel{>}{\sim}}
\def\be{\begin{equation}}  
\def\ee{\end{equation}}  
\def\bea{\begin{eqnarray}}  
\def\eea{\end{eqnarray}}  

\title{Prospects for Beyond the Standard Model Physics\\
at the Start of the LHC13 era}

\ShortTitle{BSM in LHC13 era}

\author{\speaker{Howard Baer}\\
        Homer L. Dodge Dep't of Physics and Astronomy, University of Oklahoma, Norman, OK 73019\\
        E-mail: \email{baer@nhn.ou.edu}}

\abstract{The discovery of the Higgs boson along with the 
vigorous confirmation of the SM coupled with non-appearance of new (SUSY) 
particles has led to an apparent naturalness crisis. 
We argue the crisis stems from overestimates of finetuning. A proper
evaluation of naturalness in SUSY leads to light higgsinos with mass
$\mu\sim 100-300$ GeV while other sparticles may inhabit the multi-TeV range
at little cost to naturalness. Special signatures like dilepton mass edges
in gluino pair events and same-sign dibosons from wino pairs 
might occur at LHC13. An ILC with $\sqrt{s}>2\mu$ would make the 
definitive test of naturalness in the MSSM. 
Naturalness in the QCD sector requires the axion
leading to mixed axion-WIMP dark matter. 
There appears a deep connection between electroweak naturalness and a
DFSZ-like SUSY axion sector which gives rise to the $\mu\sim f_a^2/M_P$ 
parameter.
Ultimately we expect 
detection of a higgsino-like WIMP at ton-scale DM detectors and
detection of a DFSZ-type axion at ADMX.
}

\FullConference{XXIII International Workshop on Deep-Inelastic Scattering\\
		27 April - May 1 2015\\
		Dallas, Texas}

\begin{document}

\section{Introduction} 
The LHC8 era (LHC running with $\sqrt{s}=8$ TeV) has been a grand success.\cite{Ellis:2015oda} 
Aside from a few (to be expected) minor anomalies, the Standard Model (SM) 
has been vigorously confirmed in both the electroweak and QCD sectors. In particular, a very SM-like 
Higgs boson has been discovered with mass $m_h=125.09\pm 0.24$ GeV with spin-parity given by $J^P=0^+$ and 
with very SM-like production cross sections and couplings to its various decay products.\cite{atlas_h,cms_h} 
The LHC13 era has begun with at present nearly an fb$^{-1}$ of integrated luminosity being recorded.

While the agreement of SM theory with data is generally excellent, several long-standing puzzles yet remain
that seem to require the existence of new physics.
These include:
\begin{itemize}
\item Why do neutrinos oscillate? Why are their masses so light, well below the eV scale?\cite{nureview}
\item Why is the higgs so light? Due to quadratic divergences, one expects its mass to blow up to the largest
mass scale in the theory. This usually means far beyond weak scale values.\cite{natreview}
\item Why is there no $CP$ violation in the QCD sector?\cite{axreview} 
Such $CP$ violation is to be expected from 'tHooft's solution
to the $U(1)_A$ problem via instantons and the $\theta$-vacuum. 
Yet none is seen where it is to be expected: in the neutron EDM.
\item What forms the dark matter which is pervasive throughout the universe?\cite{dmreview}
\item How did the baryon asymmetry come to be?\cite{barreview}
\end{itemize}

The first of these problems is elegantly solved by the (type-I) see-saw mechanism.\cite{seesaw} 
In this case, one introduces heavy gauge singlet right hand neutrino states $\nu_R$ into the theory with couplings
\be
{\cal L}\ni -\lambda_\nu \bar{L}_a\cdot\tilde{\phi}^a\nu_R +h.c.
\ee
where $\lambda_\nu$ is the neutrino Yukawa coupling and $L_a$ is the lepton doublet with $SU(2)$ index $a$ 
and $\tilde{\phi}=\epsilon^{ab}\phi_b^\dagger$ where $\phi_b$ is the Higgs doublet and where generation indices are suppressed.
When $\phi_b=\left({\phi^+\atop \phi^0}\right)\rightarrow\left({0\atop{v/\sqrt{2}}}\right)$ in the unitary gauge
then we develop Dirac masses $m_D=\lambda_\nu v/\sqrt{2}$ for the neutrinos. Using the conjugation property
$\psi^c\equiv C\bar{\psi}^T$ we can create {\it Majorana} neutrino spinors $\nu\equiv\nu_L+\nu_L^c$ and $N=\nu_R+\nu_R^c$
with mass matrix
\be
{\cal M}_\nu\equiv\left[\begin{array}{cc} 0 & m_D\\ m_D & M\end{array}\right]
\ee
which has eigenvalues $\sim m_D^2/M$ and $\sim M$. 
For $\lambda_\nu\simeq \lambda_t$ (as expected in various GUT theories)
then one obtains $m_{\nu 3}\sim 0.05$ eV for $M\sim 10^{15}$ GeV. This is the famous seesaw mechanism which at once explains
why the left-neutrino masses are so tiny and why the right neutrinos have not been observed. It fits exceedingly well with
GUT theories especially $SO(10)$ where the $\nu_R$ occupies the final element of the 16-dimensional spinor irrep
which contains all matter states of a single generation. While there exist other possibilities for neutrino masses
(just Dirac masses, type-II and type-III see-saw, $\cdots$) the simplicity and elegance of the type-I see-saw makes it 
hard to believe that this is not the way nature works.

\section{The Higgs mass,  naturalness and supersymmetry}

A problem now occurs. Even if one forfeits obeisance to the story of runaway quadratic divergences to the Higgs mass-- 
for instance by adopting dimensional regularization in the SM so that quadratic scalar divergences are never revealed--
they nonetheless re-appear once the neutrino see-saw is invoked. In this case, the Higgs couples to the neutrino
sector via the neutrino Yukawa coupling leading to mass corrections $m_h^2\sim (\lambda_\nu M)^2/16\pi^2$ which necessitates a
fine-tuning of order $m_h^2/M^2$ (one part in $10^{26}$) to maintain the measured value of $m_h$.\cite{Vissani:1997ys} 
The Higgs mass wants to blow up to the largest mass scale in the theory.
The well-established way to tame it is by invoking {\it supersymmetry}.\cite{susy}
In this case, the quadratic divergences necessarily cancel at all orders in perturbation theory.
The price to pay for phenomenological viability is the seeming necessity of superpartners at or around the weak scale 
(weak scale supersymmetry).\cite{wss}

It has long been the lore that weak scale SUSY requires superpartners to exist nearby to the weak scale as 
typified by $m_{W,Z,h}\sim 100$ GeV.\cite{bg,ac,dg} 
However, the failure of new physics to appear at LHC8 has raised questions
as to the nature of and existence of weak scale SUSY.\cite{doubt} 
This is further exacerbated by the rather large value of the
Higgs mass $m_h\simeq 125$ GeV which requires TeV-scale highly mixed top squarks in apparent contradiction to
expectations from naturalness.\cite{nat} 
The situation has called into question whether naturalness is indeed a guide to new physics, and
whether to abandon SUSY and search for new means to implement naturalness.\cite{alt,Barbieri:2009ev,natreview}
A whole new program of ``neutral naturalness''
has emerged which seeks to cancel offending divergences by appealing to hidden sector states which are neutral
under the SM charges\cite{neut,craig,neutsig}. 
Indeed, some authors go so far as to proclaim a {\it crisis} for physics\cite{lykken}.

Before jumping ship prematurely, it pays to scrutinize more quantitatively the notion of naturalness 
for weak scale SUSY (WSS). 
After all, experiment has had three chances to disprove WSS and each time SUSY has met the challenge.
As a reminder, 
1. precision measurement of gauge couplings at LEP are in accord with SUSY gauge coupling unification,\cite{gauge} 
2. the measured value of the top quark mass was found to be in accord with that required by SUSY for successful 
radiatively-driven electroweak symmetry breaking\cite{rewsb} and 
3. the measured value of $m_h$ falls squarely within the range required by
SUSY where $m_h\alt 135$ GeV is required in the MSSM\cite{mhiggs}.

\subsection{Electroweak naturalness}

While SUSY provides a solution to the big hierarchy problem via cancellation of quadratic divergences, 
the increasingly severe sparticle mass bounds have accentuated a growing Little Hierarchy: why are $m_{W,Z,h}\ll m_{sparticle}$
when $m_{sparticle}$ helps determine $m_{weak}$? The most direct connection comes from the MSSM 
scalar potential minimization conditions where it is found that
\bea
\frac{m_Z^2}{2} &=& \frac{(m_{H_d}^2+\Sigma_d^d)-(m_{H_u}^2+\Sigma_u^u)\tan^2\beta}{(\tan^2\beta -1)}
-\mu^2\\
&\simeq &-m_{H_u}^2-\mu^2-\Sigma_u^u
\label{eq:mzs}
\eea
where $m_{H_u}^2$ and $m_{H_d}^2$ are the {\it weak scale} soft SUSY breaking Higgs masses, $\mu$
is the {\it supersymmetric} higgsino mass term and $\Sigma_u^u$ and $\Sigma_d^d$ contain
an assortment of loop corrections to the effective potential. 
From this equation, naturalness is ensured if each term on the right-hand-side is comparable to $m_Z^2/2$.
A naturalness measure $\Delta_{EW}$ has been introduced which
compares the largest contribution on the right-hand-side of Eq. \ref{eq:mzs} 
to the value of $m_Z^2/2$. If they are comparable ($\Delta_{EW}\alt 10-30$), 
then no unnatural fine-tunings are required to generate $m_Z=91.2$ GeV. 
The main requirement is then that
$|\mu |\sim m_Z$\cite{Chan:1997bi,Barbieri:2009ev,hgsno} (with $\mu \agt 100$ GeV to accommodate LEP2 limits 
from chargino pair production searches) 
and also that $m_{H_u}^2$ is driven radiatively to small, 
and not large, negative values~\cite{ltr,rns}.\footnote{Some recent work on theories with naturalness and
heavy higgsinos include \cite{Cohen:2015ala,Nelson:2015cea,spmartin}. 
Such theories tend to introduce unwanted exotica such as adjoint scalars.}
Also, the top squark contributions to the radiative corrections $\Sigma_u^u(\tst_{1,2})$ 
are minimized for TeV-scale highly mixed top squarks\cite{ltr}. This latter condition  also lifts 
the Higgs mass  to $m_h\sim 125$ GeV.
The measure $\Delta_{EW}$ is pre-programmed in the Isasugra SUSY spectrum generator\cite{isajet}.

\subsection{Large logs}

In contrast to the above, one often instead hears that the log corrections to $m_h^2$ are too large in SUSY:
\be
\delta m_h^2\sim -\frac{3f_t^2}{16\pi^2}m_{\tilde t}^2\log\left(\Lambda^2/m_{\tilde t}^2\right) .
\ee
Taking $\Lambda$ as high as $m_{GUT}\sim 2\times 10^{16}$ GeV and $m_{\tilde t}\sim 5-10$ TeV, then indeed
one expects $\sim 0.1-1\%$ fine-tuning. 

We feel this approach is too simplistic and leads to overestimates in EW finetuning for SUSY.\cite{comp} 
To avoid overestimates, we have stated the fine-tuning rule:\cite{seige}
\begin{quotation}
When evaluating fine-tuning, it is not permissible to claim fine-tuning 
of {\it dependent} quantities one against another.
\end{quotation}
In the above large log case, the Higgs mass is given by
\be
m_h^2\simeq \mu^2+m_{H_u}^2(\Lambda) +\delta m_{H_u}^2
\ee
where the logs enter $\delta m_{H_u}^2$ properly via the RGE
\be
\frac{dm_{H_u}^2}{dt}=\frac{1}{8\pi^2}\left(-\frac{3}{5}g_1^2M_1^2-3g_2^2M_2^2+\frac{3}{10}g_1^2 S+3f_t^2 X_t\right)
\label{eq:mHu}
\ee
where $t=\ln (Q^2/Q_0^2)$,
$S=m_{H_u}^2-m_{H_d}^2+Tr\left[{\bf m}_Q^2-{\bf m}_L^2-2{\bf m}_U^2+{\bf m}_D^2+{\bf m}_E^2\right]$
and $X_t=m_{Q_3}^2+m_{U_3}^2+m_{H_u}^2+A_t^2$.
The above large log can be found by neglecting gauge terms and $S$ ($S=0$ in models with scalar soft term universality 
but can be large in models with non-universality),
and also neglecting the $m_{H_u}^2$ contribution to $X_t$ and the fact that $f_t$ and the soft terms
evolve under $Q^2$ variation. 
Especially egregious is the neglect of $m_{H_u}^2$. 
The more one increases $m_{H_u}^2(\Lambda )$, then the greater is
the cancelling correction $\delta m_{H_u}^2$. (This is different from the SM case where the leading divergences
and the tree level Higgs mass are independent.)
By collecting dependent contributions, instead one requires each of the two terms on the RHS of
\be
m_h^2\simeq \mu^2+\left( m_{H_u}^2(\Lambda )+\delta m_{H_u}^2\right)
\ee
to be comparable to $m_h^2$. This then is the same requirement as for low $\Delta_{EW}$ since
$m_{H_u}^2(weak)=m_{H_u}^2(\Lambda )+\delta m_{H_u}^2$. 

\subsection{BG fine-tuning}

Traditionally, naturalness is quantified via the EENZ/BG measure\cite{eenz,bg} 
\be 
\Delta_{BG}\equiv max_i \left| \frac{\partial m_Z^2}{\partial p_i}\right|
\ee
where $p_i$ are the fundamental parameters of the theory. Using $p_i$ as the weak scale values 
$\mu^2,\ m_{H_u}^2$ etc. as in the pMSSM, then $\Delta_{BG}\sim \Delta_{EW}$. But if we evaluate
$\mu^2$ and $m_{H_u}^2$ in terms of GUT scale parameters as in gravity-mediation, then instead we have
(for the case of $\tan\beta =10$)~\cite{abe,martin,feng}
\be
m_Z^2 \simeq  -2.18\mu^2 + 3.84 M_3^2-0.65 M_3A_t-1.27 m_{H_u}^2 -0.053 m_{H_d}^2
+0.73 m_{Q_3}^2+0.57 m_{U_3}^2 +\cdots .
\label{eq:mzsHS}
\ee
With $M_3\simeq m_{\tg}$ and $m_{\tg}\agt 1300$ GeV from LHC8 searches, we again find WSS to be highly tuned
with $\Delta_{BG}\sim 800$.

However, in supergravity theory it is something of a theorem that for any particular hidden sector the
high-scale soft terms are all calculable as multiples of the gravitino
mass $m_{3/2}$\cite{sw}. If we vary $m_{3/2}$, the soft terms all vary
accordingly: {\it i.e.} they are {\it not independent} in SUGRA models. 
The soft terms are only independent in the effective theories where soft terms are introduced to 
parameterize our ignorance of the SUSY breaking sector.
By combining the dependent soft SUSY breaking terms, 
then the $Z$ mass can be expressed as\cite{seige} 
\be
m_Z^2\simeq -2\mu^2 (\Lambda )-a m_{3/2}^2 ~,
\label{eq:mzsm32}
\ee
with $a$ being a certain proportionality factor dependent on each 
soft mass spectrum. Using Eq.~\ref{eq:mzsm32}-- and since $\mu$ hardly
evolves from $\Lambda$ to $m_{\rm weak}$-- we have $am_{3/2}^2\simeq 2
m_{H_u}^2 ({\rm weak})$. Even if $m_{3/2}$ is large (as implied by LHC8 limits
for gravity-mediation), then one may still generate natural models if
the coefficient $a$ is small. Under the combination of dependent soft
SUSY breaking terms, then low $\Delta_{\rm BG}$ implies the same as
low $\Delta_{EW}$: that $\mu\sim m_{\rm weak}$ and that $m_{H_u}^2$ is driven to
small and not large negative values. 

\section{Where is SUSY?}

\begin{figure}[tbp]
\begin{center}
\includegraphics[height=0.4\textheight]{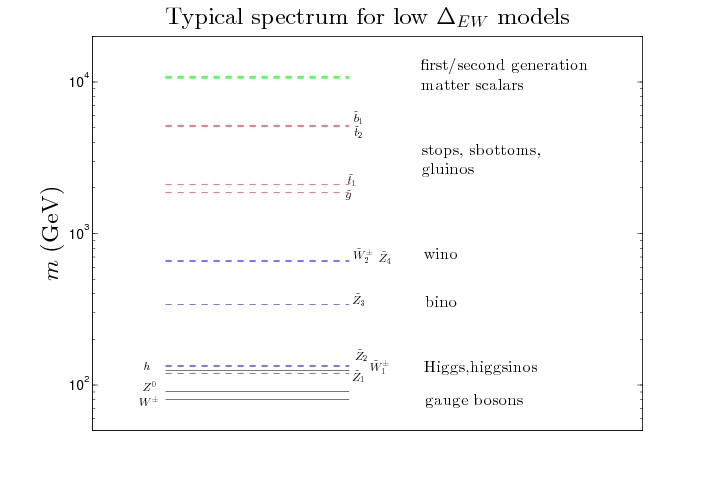}
\caption{
Typical sparticle mass spectrum from SUSY models with low $\Delta_{EW}$,
{\it i.e.} radiatively-driven naturalness.
\label{fig:spec}}
\end{center}
\end{figure}

By properly evaluating naturalness, we find that $\mu^2$ and $m_{H_u}^2$ are $\sim m_{weak}$ while
the sparticles can be much heavier, of order $m_{3/2}$. The Little Hierarchy $\mu\ll m_{3/2}$ is totally
acceptable. When we say SUSY particles should be $\sim m_{weak}$, it is really only the higgsinos 
$\tw_1^\pm , \tz_{1,2}$ which are required to have mass $\sim \mu\sim m_{weak}$. Since the higgsinos
are rather compressed-- detailed evaluations show that the interhiggsino mass gap is typically 10-30 GeV\cite{rns}--
then the heavier higgsino decay products are quite soft and buried beneath a
huge background of soft QCD events at LHC. The lightest higgsino is the LSP and escapes detection.
A typical sparticle mass spectrum with radiatively-driven naturalness is shown in Fig. \ref{fig:spec}.

\begin{figure}[tbp]
\begin{center}
\includegraphics[height=0.3\textheight]{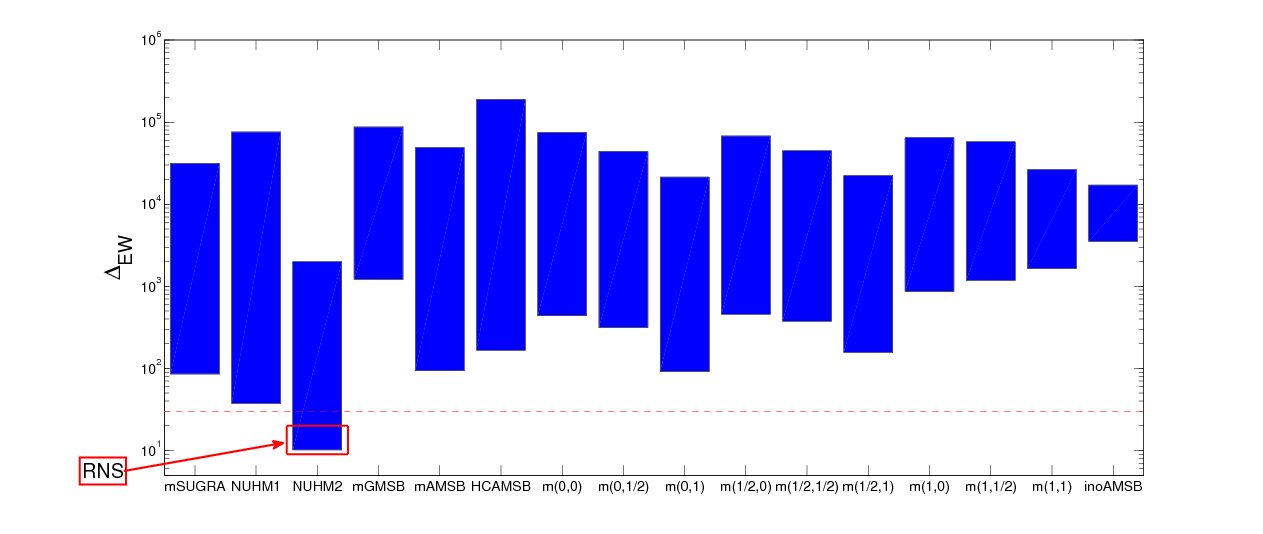}
\caption{
Plot of $\Delta_{EW}$ from a scan over parameter space for 16 different SUSY models.
Naturalness requires $\Delta_{EW}\alt 30$. 
Only the RNS portion of the NUHM2 model survives.
\label{fig:hist}}
\end{center}
\end{figure}

The measure $\Delta_{EW}$ can be used to rule out models based on whether or not they are natural (since unnatural
models are almost assuredly wrong models). A sampling of 16 models is shown in Fig. \ref{fig:hist}
taken from Ref. \cite{seige}.
Requiring $\Delta_{EW}\alt 30$ and $m_h:123-127$ GeV rules out all models (including the highly popular mSUGRA/CMSSM model)
except for one region of parameter space of the two-extra-parameter non-universal Higgs model (NUHM2) 
where radiatively-driven natural SUSY (RNS) occurs.

Upper bounds on sparticle masses versus $\Delta_{EW}\simeq \Delta_{BG}$ are evaluated in Ref's \cite{rns,upper}.
For $\Delta_{EW}\alt 10$, then $m_{\tg}\alt 2$ TeV (within LHC13 range at high luminosity), $\mu <200$ GeV 
(within range of ILC500) and $m_{\tst_1}\alt 1.5$ TeV (perhaps beyond LHC13 reach). For $\Delta_{EW}\alt 30$
which we regard as a conservative upper limit, then $m_{\tg}\alt 4$ TeV, $\mu\alt 350$ GeV and $m_{\tst_1}\alt 3$ TeV.

\section{Radiatively-driven natural SUSY at colliders}

\begin{figure}[!htb]
\begin{center}
\includegraphics[width=110mm]{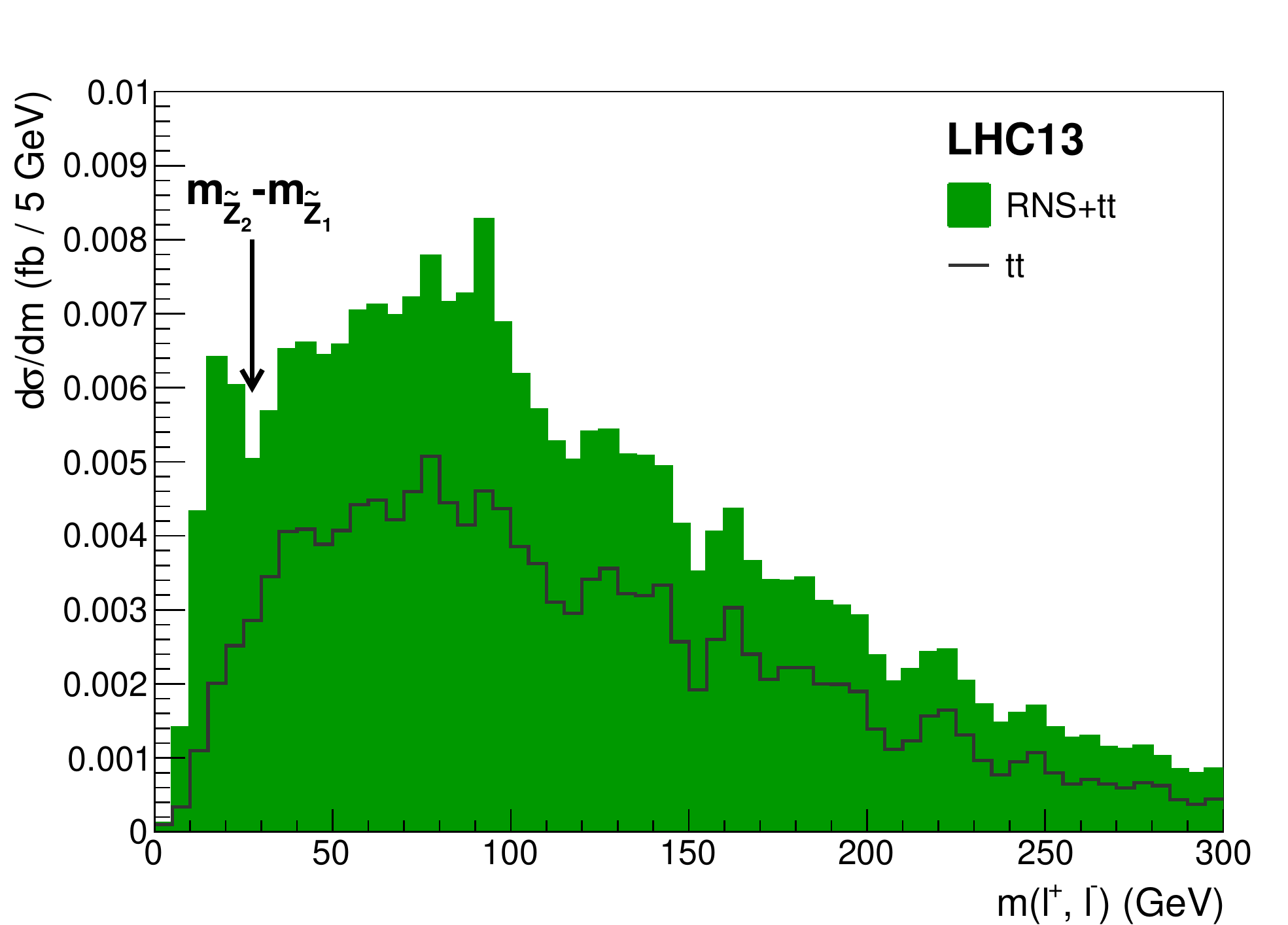}
\caption{Invariant mass of OSSF leptons. The dilepton mass edge and the $Z$ peak
are visible for the RNS model. We require $n(\textit{b-jets}) \geq 3$.}
\label{fig:invmassRNS}
\end{center}
\end{figure}

Thus, LHC13 with $300-1000$ fb$^{-1}$ of integrated luminosity will explore up to $\Delta_{EW}\sim 10$ via gluino pair 
production. These events, if seen, should be characteristic\cite{lhc,LSPtypes}: rich in $b$- and top-jets while the gluino cascade
decays should include a characteristic dilepton mass edge $m(\ell^+\ell^- )<m_{\tz_2}-m_{\tz_1}\sim 10-30$ GeV:
see Fig. \ref{fig:invmassRNS}.

In addition, a qualitatively new SUSY signal emerges. For natural SUSY spectra, then wino pair production
in the form $pp\to \tw_2^\pm\tz_4$ production is the dominant visible cross section. The winos decay as
$\tw_2\to W\tz_1$ and $\tz_4\to W^\pm\tw_1^\mp$. Thus, half the wino pair production events contain same-sign diboson
pairs which are distinct from usual SS dilepton events in that they should be relatively jet-free.
SM backgrounds to the SSdB signature are tiny.
The reach via SSdB signal exceeds that from $\tg\tg$ production for higher integrated luminosity values\cite{lhcltr,lhc}.
\begin{figure}[tbp]
\begin{center}
\includegraphics[height=0.2\textheight]{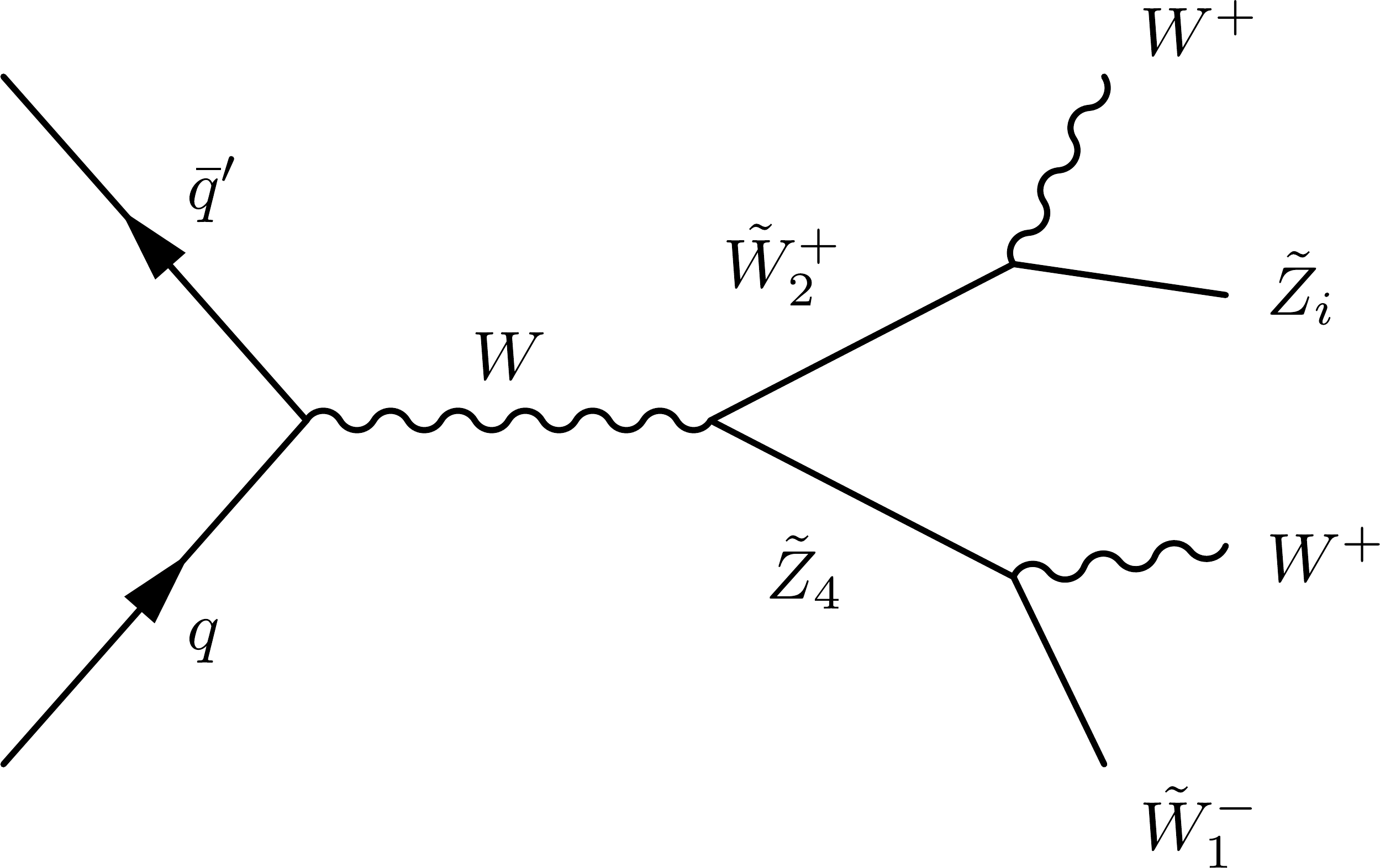}
\caption{Diagram depicting same-sign diboson production at LHC in SUSY models 
with light higgsinos.
\label{fig:diagram}}
\end{center}
\end{figure}

\begin{figure}[tbp]
\begin{center}
\includegraphics[width=9cm,clip]{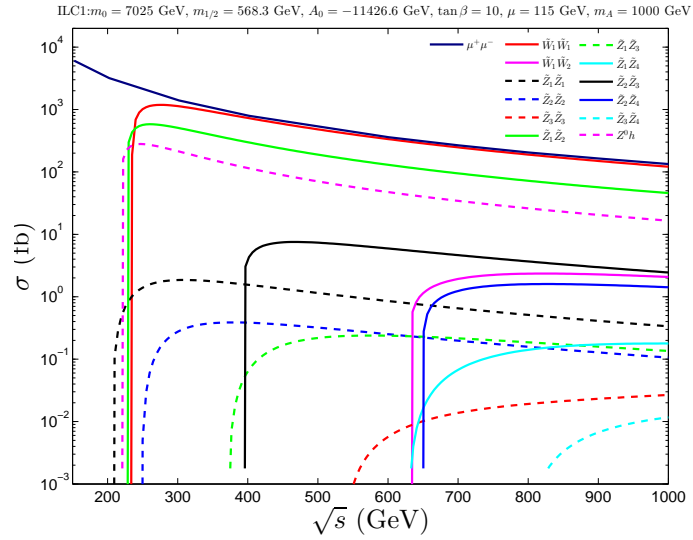}
\caption{Sparticle production cross sections vs. $\sqrt{s}$ for
  unpolarized beams at an $e^+e^-$ collider for the ILC1 benchmark point
  listed in Ref. \cite{ilc}.}
\label{fig:xsec}
\end{center}
\end{figure}

The really neat signature of natural SUSY is higgsino pair production at ILC. In this case, SUSY can be easily discovered
and precision measurements verifying the higgsino nature and testing gaugino mass unification can be made. 
The ILC will be a higgsino factory in addition to a Higgs factory\cite{ilc}: see Fig. \ref{fig:xsec}.

\section{Naturalness in QCD: the need for axions}
\label{sec:qcd}

If we insist on naturalness in the electroweak sector, then it is only fair to 
insist as well on naturalness in the QCD sector.
In the early days of QCD, it was a mystery why the two-light-quark chiral symmetry U(2)$_L\times$U(2)$_R$
gave rise to three and not four light pions~\cite{U1}. 
The mystery was resolved by 't Hooft's discovery of the QCD theta vacuum which didn't respect 
the remnant $U(1)_A$ symmetry~\cite{tHooft}. 
As a consequence of the theta vacuum, one expects the presence of a term 
\be
{\cal L}\ni \frac{\bar{\theta} g_s^2}{32\pi^2}F_{A\mu\nu}\tilde{F}_A^{\mu\nu}
\ee 
in the QCD Lagrangian (where $\bar{\theta}=\theta+\arg(\det({\cal M}))$ and ${\cal M}$ 
is the quark mass matrix). Measurements of the neutron EDM constrain $\bar{\theta}\lesssim 10^{-10}$ 
leading to an enormous fine-tuning in $\bar{\theta}$: the so-called strong CP problem.

The strong CP problem is elegantly solved by Peccei, Quinn, Weinberg and Wilczek (PQWW)~\cite{pqww}
via the introduction of PQ symmetry and the concomitant (invisible~\cite{ksvz,dfsz}) axion: 
the offending term can dynamically settle to zero.
The axion is a valid dark matter candidate in its own right~\cite{axdm}.

Introducing the axion in a SUSY context solves the strong CP problem and 
renders naturalness to QCD. 
As a bonus, in the context of the SUSY DFSZ axion model~\cite{dfsz} 
where the Higgs superfields carry PQ charge,
one gains an elegant solution to the SUSY $\mu$ problem. 
The most parsimonius implementation of the strong CP solution
involves introducing a single MSSM singlet superfield $S$ carrying PQ charge $Q_{PQ}=-1$ while the
Higgs fields both carry $Q_{PQ}=+1$. The usual $\mu$ term is forbidden, but we have a 
superpotential~\cite{kn,susydfsz}
\be
W_{\rm DFSZ}\ni \lambda\frac{S^2}{M_P}H_uH_d .
\ee
If PQ symmetry is broken and $S$ receives a VEV $\langle S\rangle\sim f_a$, then a weak scale
$\mu$ term
\be
\mu\sim \lambda f_a^2/M_P
\ee
is induced which gives $\mu\sim m_Z$ for $f_a\sim 10^{10}$ GeV. Although Kim-Nilles sought to relate
the PQ breaking scale $f_a$ to the hidden sector mass scale 
$m_{\rm hidden}$~\cite{kn}, we see 
now that the Little Hierarchy 
\be
\mu\sim m_Z\ll m_{3/2}\sim {\rm multi-TeV} 
\ee
could emerge due to a mis-match between PQ breaking scale and hidden sector 
mass scale $f_a\ll m_{\rm hidden}$. 

An elegant model which produces the above hierarchy was proposed by 
Murayama, Suzuki and Yanagida (MSY)~\cite{msy}. In the MSY model, PQ symmetry is broken
radiatively by driving one of the PQ scalars $X$ to negative mass-squared values in 
much the same way that electroweak symmetry is broken by radiative corrections 
driving $m_{H_u}^2$ negative.
Starting with multi-TeV scalar masses, the radiatively-broken PQ symmetry induces 
a SUSY $\mu $ term $\sim 100$ GeV~\cite{radpq} while at  the same time generating 
intermediate scale Majorana masses for right-hand neutrinos.
In models such as MSY, the Little Hierarchy $\mu\ll m_{3/2}$ is no problem at all 
but is instead just a reflection of the mis-match between PQ and hidden sector mass scales.

\section{Mixed axion-WIMP dark matter}

\subsection{Dark matter: an axion/WIMP admixture?}

As mentioned above, to allow for both electroweak and QCD naturalness, one needs
a model including both axions and SUSY. In such a case, the axion field is promoted to a superfield
which contains a spin-0 $R$-parity even saxion $s$ and a spin-$1/2$ R-parity odd axino $\ta$.
Typically in SUGRA one expects the saxion mass $m_s\sim m_{3/2}$ and the axino mass $m_{\ta}\alt m_{3/2}$.\cite{axmass}
The dark matter  is then comprised of two particles: the axion along with the LSP 
which is a higgsino-like WIMP.
This is good news for natural SUSY since thermal higgsino-like WIMPs are typically underproduced
by a factor 10-15 below the measured dark matter abundance. The remainder can be comprised of axions.
 
\begin{figure}[tbp]
\begin{center}
\includegraphics[height=0.3\textheight]{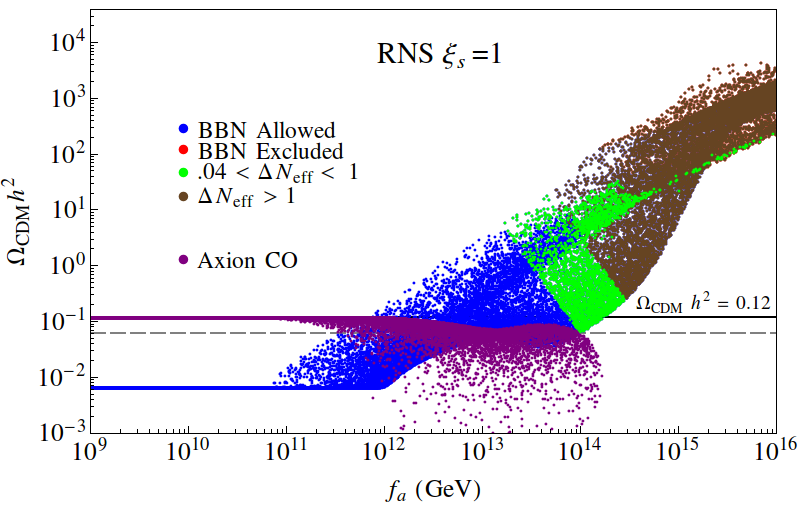}
\caption{
The neutralino relic density from a scan over
SUSY DFSZ parameter space for the RNS benchmark case 
labelled SUA with $\xi=1$.
The grey dashed line shows the points where DM consists of 50\% axions and
50\% neutralinos.
The red BBN-forbidden points occur at $f_a\agt 10^{14}$ GeV and are covered over by the brown 
$\Delta N_{eff}>1.6$ coloration. 
This latter region is excluded by Planck limits\cite{Planck:2015xua} of dark radiation
as parametrized by additional neutrino species beyond the SM value.
\label{fig:sua1}}
\end{center}
\end{figure}

The amount of dark matter generated in the early universe depends sensitively on the
properties of the axino and the saxion in addition to the SUSY spectrum and the axion.
For instance, thermally produced axinos can decay into LSPs after neutralino freeze-out 
thus augmenting the LSP abundance\cite{az1}. 
If too many WIMPs are produced from axino decay, then they may re-annihilate at the axino decay temperature\cite{az1}.
Saxions can be produced thermally or via coherent oscillations
(important at large $f_a$) and their decays can add to the LSP abundance, produce extra dark radiation
in the form of axions or dilute all relics via entropy production from decays to SM particles\cite{bbl}.
The calculation of the mixed axion-WIMP abundance requires solution of eight coupled Boltzmann
equations. Results from a mixed axion-higgsino dark matter calculation in natural SUSY are
shown in Fig. \ref{fig:sua1}\cite{dfsz2}. At low $f_a\sim 10^{10}$ GeV, then the thermal value of WIMP 
production is maintained since axinos decay before freeze-out. In this case the DM is axion-dominated\cite{bbc}.
For higher $f_a$ values, then axinos and saxions decay after freeze-out thus augmenting the 
WIMP abundance. For very large $f_a\agt 10^{14}$ GeV, then WIMPs are overproduced and those cases would be
excluded. Many of the high $f_a$ models are also excluded via violations of BBN constraints and by
overproduction of dark radiation- as parametrized by the effective number of extra neutrinos
in the universe $\Delta N_{eff}$.

\begin{figure}[tbp]
\begin{center}
\includegraphics[height=0.4\textheight]{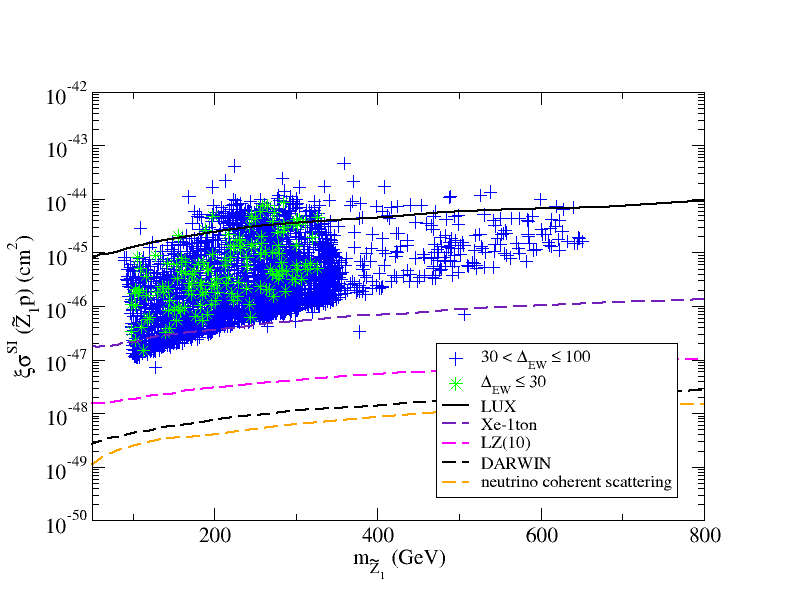}
\caption{
Plot of rescaled higgsino-like WIMP spin-independent 
direct detection rate $\xi \sigma^{SI}(\tz_1 p)$ 
versus $m(higgsino)$ from a scan over NUHM2 parameter space with $\Delta_{EW}<30$ (green stars)
and $\Delta_{EW}<100$ (blue crosses). 
We also show the current reach of LUX and projected reaches of 
several ton-scale WIMP detectors.
Plot from Ref. \cite{rnsdm}.
\label{fig:SI}}
\end{center}
\end{figure}

As far as dark matter detection goes, WIMP production in RNS was examined in Ref. \cite{bbm}.
There, it is emphasized that the relevant theory prediction for WIMP direct detection is the
quantity $\xi\sigma^{SI} (\tz_1 p)$ where $\xi =\Omega_{\tz_1}h^2/0.12$ to reflect the
possibility that the WIMP local abundance may be highly depleted, and perhaps axion-dominated.
Nonetheless, WIMPs should be ultimately detected by ton-scale noble liquid detectors
because naturalness insures that the WIMP-Higgs coupling-- which is a product of 
higgsino and gaugino components-- is never small (see Fig. \ref{fig:SI}).
Prospects for indirect detection of higgsino-like WIMPs from WIMP-WIMP annilations 
to gamma rays or anti-matter are less lucrative since then the expected detection rates must
be scaled by $\xi^2$.

\begin{figure}
\begin{center}
\includegraphics[height=4.6cm]{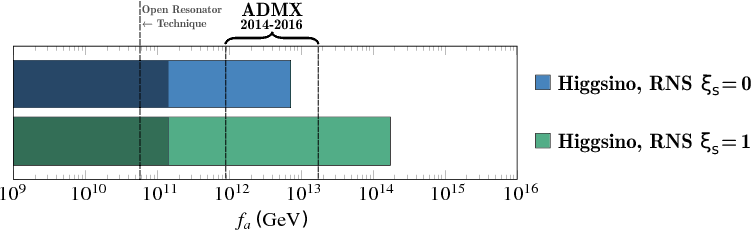}
\caption{Range of $f_a$ which is allowed in each PQMSSM scenario for the 
RNS benchmark models. 
Shaded regions indicate the range of $f_a$ where $\theta_i>3$.
\label{fig:bar}}
\end{center}
\end{figure}

Meanwhile, we would also expect ultimate detection of axions if natural SUSY prevails\cite{axsearch}.
In Fig.~\ref{fig:bar} we display the range of $f_a$ where valid solutions for the relic abundance 
of mixed axion-higgsino CDM can be found in SUSy models with radiatively driven naturalness.
The upper bar shows the range of $f_a$ for $\xi_s=0$ (no saxion-axion coupling) while the
lower bar shows the range for $\xi_s=1$ (axion-saxion coupling turned on). 
The darker shaded parts of the bars denote 
$\theta_i$ values $>3$ which might be considered less plausible or fine-tuned.
We also show by the bracket the range of $f_a$, assuming the bulk of DM is axion, 
which is expected to be probed by the ADMX experiment within the next several years. 
This region probes the most natural region
where $\theta_i\sim 1$. We also show a further region of lower $f_a$ which might be explored by
a new open resonator technology. About a decade of natural $f_a\sim 10^{14}$ GeV 
seems able to elude ADMX searches for the $\xi_s=1$ case.%

\section{Conclusions}

Our conclusions are summarized as follows:
\begin{itemize}
\item There is no naturalness crisis for SUSY even though many popular 
models such as CMSSM/mSUGRA are ruled out by
naturalness\cite{seige}. 
By requiring naturalness, we are pushed into a very specific mass spectrum charcterized by light higgsinos
with mass $\sim 100-300$ GeV while other sparticles lie typically in the 2-20 TeV range at little cost to naturalness.
In particular, gluinos may range up to 2 (4) TeV for $\Delta_{EW}\alt 10\ (30)$. 
Stops typically lie in the several TeV range.
\item LHC will probe the most lucrative region of natural SUSY parameter space via $\tg\tg$ production and ultimately same-sign diboson production. Gluino pair events should contain a characteristic OS/SF dilepton mass edge $\alt 10-30$ GeV.
\item Natural SUSY gives {\it tremendous} motivation to build ILC. For natural SUSY, ILC will be a higgsino factory
for $\sqrt{s}>2\mu$. 
\item Requiring naturalness in the QCD sector requires axions and the SUSY DFSZ axion model leads to a 
natural solution to the SUSY mu problem where $\mu\ll m_{3/2}$. Then we expect mixed axion-higgsino-like-WIMP
dark matter.
\item Even with a suppressed local abundance of higgsino-like WIMPs, a  signal should be seen at ton-scale
noble liquid detectors. We also expect an axion signal to emerge at ADMX although the axion mass may lie
somewhat above the projected ADMX search region.
\end{itemize}

{\bf Acknowledgments:} I would like to thank the organizers of DIS2015 for 
their kind invitation to speak. I thank my many estimable collaborators.
This work was supported in part by the 
US Department of Energy Office of High Energy Physics.

\end{document}